\documentclass[a4paper,12pt]{article}
\pdfoutput=1 

\usepackage{jcappub} 
\usepackage[utf8]{inputenc}
\usepackage{graphics}
\usepackage{epstopdf}
\usepackage{hyperref}
\usepackage{amsmath}
\usepackage{bigints}
\usepackage{relsize}
\usepackage{psfrag}
\usepackage{epsfig}
\usepackage{subfig}
\usepackage{xcolor}
\usepackage{url}

\usepackage[T1]{fontenc} 
\usepackage[percent]{overpic}
\usepackage{breqn}

 \normalsize
\newcommand{\bmat}{\left(\begin{array}}
\newcommand{\emat}{\end{array}\right)}
\newcommand{\be}{\begin{equation}}
\newcommand{\ee}{\end{equation}}
\newcommand{\bea}{\begin{eqnarray}}
\newcommand{\eea}{\end{eqnarray}}

\usepackage{booktabs}
\usepackage{siunitx}
\def\lsim{\raise0.3ex\hbox{$\;<$\kern-0.75em\raise-1.1ex\hbox{$\sim\;$}}}
\def\gsim{\raise0.3ex\hbox{$\;>$\kern-0.75em\raise-1.1ex\hbox{$\sim\;$}}}

\title{\boldmath Primordial monopoles, black holes and gravitational waves}

\author[a]{Ahmad Moursy,}
\author[b]{Qaisar Shafi}

\affiliation[a]{Department of Basic Sciences, Faculty of Computers and Artificial Intelligence,  \\
Cairo University, Giza 12613, Egypt}
\affiliation[b]{Bartol Research Institute, Department of Physics and Astronomy, \\
	University of Delaware, Newark, DE 19716, USA}
\emailAdd{a.moursy@fci-cu.edu.eg}
\emailAdd{qshafi@udel.edu}
\abstract{We show how topologically stable superheavy magnetic monopoles and primordial black holes can be generated at observable levels by the waterfall field in hybrid inflation models based on grand unified theories. In $SU(5) \times U(1)_\chi$ grand unification, the monopole mass is of order $4 \times 10^{17}$ GeV, and it carries a single unit ($2 \pi /e$) of Dirac magnetic charge as well as screened color magnetic charge. The monopole density is partially diluted to an observable value, and accompanied with the production of primordial black holes with mass of order $10^{17}$-$10^{19}$ g which may make up the entire dark matter in the universe. The tensor to scalar ratio $r$ is predicted to be of order $10^{-5}$ - $10^{-4}$ which should be testable in the next generation of CMB experiments such as CMB-S4 and LiteBIRD. The gravitational wave spectrum generated during the waterfall transition is also presented. The observed baryon asymmetry can be explained via leptogenesis.
}
\begin{document}
\maketitle
\flushbottom
\section{Introduction}
\label{sec:intro}
﻿Grand Unified Theories (GUTs) based on $SU(5)$ \cite{Georgi:1974sy} and $SO(10)$~\cite{Georgi:1974my,Fritzsch:1974nn,Lazarides:1980cc} predict the existence of a superheavy topologically stable magnetic monopole that carries a single quantum ($2 \pi /e$) of Dirac magnetic charge as well as color magnetic field which is screened \cite{Lazarides:2023iim}. The mass of this monopole is about ten times larger than the GUT scale, namely of order $10^{17}$ GeV or so.\footnote{This is analogous to the monopole mass of order $10 \times M_W$, as originally shown by ‘t Hooft and Polyakov \cite{Polyakov:1974ek} in  a toy model based on $SO(3)$ \cite{tHooft:1974kcl}.} 

GUTs based on $SO(10)$ also predict the presence of topologically stable strings if the symmetry breaking is implemented with VEVs arising exclusively from scalar fields in tensor representations \cite{Kibble:1982ae}. An unbroken $Z_2$ gauge symmetry is responsible for the appearance of these strings, and their mass per unit length is determined by the symmetry breaking scale associated with the appearance of this $Z_2$ symmetry. Among other things, this $Z_2$ symmetry plays the role of “matter” parity in supersymmetric models, and it is also used to provide stable dark matter particle in non-supersymmetric models~\cite{Mambrini:2015vna,Boucenna:2015sdg,Nagata:2015dma,Nagata:2016knk,Ferrari:2018rey}. 

With the advent of inflationary cosmology \cite{Linde:1993cn,Dvali:1994ms} it has been a challenge to show how topological defects, especially the  superheavy ones, can survive primordial inflation. An early example based on $SU(5)$ inflation \cite{Shafi:1983bd,Antusch:2023mxx} shows how superheavy global strings survive inflation \cite{Shafi:1984tt}. Subsequently, it was shown how one can realize an observable number density of intermediate scale magnetic monopoles in realistic inflationary models based on $SO(10)$~\cite{Senoguz:2015lba,Chakrabortty:2020otp,Maji:2022jzu}.

In a recent paper \cite{Lazarides:2023rqf} based on flipped $SU(5)$ and hybrid inflation, it was shown how one can realize a metastable cosmic string scenario \cite{Lazarides:2022jgr,Lazarides:2023ksx,Afzal:2023kqs,King:2023wkm,Buchmuller:2023aus,Lazarides:2024niy,Roshan:2024qnv} with a dimensionless string tension parameter $G\mu \sim 10^{-6}$. The challenge here was to inflate away the superheavy monopoles entirely, but not the associated cosmic strings that are nearly just as heavy. The latter experience a limited amount of inflation and emit gravitational waves after they re-enter the horizon. The strings are metastable and eventually decay from the quantum tunneling of the monopoles.

Primordial Black Holes (PBHs) may exist in the early universe and constitute the entire dark matter~\cite{Carr:1974nx,Carr:1975qj,Chapline:1975ojl}. Inflationary scenarios such as hybrid inflation~\cite{Linde:1993cn,Dvali:1994ms,Ibrahim:2022cqs,Lazarides:2023rqf} predict PBHs when  sufficiently large density
fluctuations collapse gravitationally in the early universe~\cite{Garcia-Bellido:1996mdl,Clesse:2015wea}. With an intermediate waterfall regime, hybrid inflation is drived by a single inflaton field at the start of inflation, with the rest of the {\it waterfall} fields frozen until the waterfall transition. During the waterfall, curvature perturbations are enhanced, resulting in a peak in the curvature power spectrum. The latter enhancement sources  scalar induced gravitational waves (SIGW) through second order effects in perturbation theory~\cite{Ananda:2006af,Baumann:2007zm,DeLuca:2020agl}. The SIGW energy density scales as $a^{-4}$, similar to the radiation energy density, with rough estimation of the ratio $\Omega_{\rm GW}h^2/ \Omega_{\rm r}h^2 \sim P_\zeta^2$, which is almost conserved until the present day \cite{Escriva:2022duf}. For Gaussian perturbations and the power spectrum peak $P_\zeta\sim 10^{-2}$, $\Omega_{\rm GW}h^2\sim 10^{-9}$.

In this paper we show how to extend the discussion in Ref.~\cite{Lazarides:2023rqf} to realistic inflationary GUT models that produce an observable number density of primordial magnetic monopoles as well as primordial black holes and scalar induced gravitational waves. In a particularly simple but realistic model based on $SU(5) \times U(1)_\chi$~~\cite{Pal:2019pqh,Ahmed:2023rky,Ahmed:2022vlc}, a maximal subgroup of $SO(10)$, the topologically stable superheavy monopole carries a single unit of Dirac magnetic charge as well as some screened color charge. We discuss how this monopole arises at an observable level from the spontaneous breaking of a waterfall field in the adjoint representation of $SU(5)$ that experiences a limited e-folds of inflation. The model also produces primordial black holes associated with the waterfall transition, which can provide the desired dark matter relic abundance. The gravitational wave spectrum induced by the enhanced scalar perturbations during the waterfall transition is also discussed. In this particular example the cosmic strings associated with the spontaneous breaking of $U(1)_\chi$ are inflated away. An extension of this scenario should yield primordial monopoles, cosmic strings and black hole dark matter with observable signatures. Interestingly, we show that the $SU(5)$ symmetry breaking scale  needed for grand unification as well as observable monopoles and PBHs as DM is of order $ 3\times 10^{16}$ GeV, which is realized if TeV scale vector-like fermions are added as shown in Ref.~\cite{Gogoladze:2010in}. The proton lifetime is estimated to be of order $10^{36}-10^{37}$ yrs~\cite{Lazarides:2023rqf}, which may prove challenging for the Hyper-Kamiokande experiment~\cite{Dealtry:2019ldr}.

The paper is organized as follows. In section~\ref{sec:model} we discuss a realistic $\chi SU(5)$ model that produces topologically stable monopoles and provide the field content. We investigate the inflationary scenario and dynamics of the scalar fields in section~\ref{sec:infdynamic}, and in section~\ref{sec:observables} compute the inflation observables. Reheating after inflation and leptogenesis are discussed in section~\ref{sec:reheat-leptogenesis}. The waterfall dynamics is studied in section~\ref{sec:wfdynamics} including generation of enhanced curvature perturbations as well as production and observability of the superheavy monopoles. We explore in section~\ref{sec:PBH-GW} the production of primordial black holes and scalar induced gravitational waves. Our conclusions are summarized in section~\ref{sec:conc}.
%
\section{  $\chi SU(5)$ monopoles}
\label{sec:model}
The model is based on the gauge symmetry $SU(5)\times U(1)_\chi$ ($\chi SU(5)$), a maximal subgroup of $SO(10)$, which is broken to the SM in two steps as follows: %
 \bea \label{eq:SBpattern} 
SU(5)\times U(1)_\chi   &\xrightarrow[]{\left<\Phi\right>} &  SU(5)  \nonumber \\
&\xrightarrow[]{\left<\Psi\right>} &  SU(3)_c \times SU(2)_L \times U(1)_{Y} .
\eea
Here  $\Phi$ is a complex scalar field that is charged under $U(1)_\chi$ and singlet under $ SU(5)$, while $\Psi \equiv \mathbf{24}_H$ is in the $ SU(5)$ adjoint representation, but is neutral under $U(1)_\chi$. In the first step in~(\ref{eq:SBpattern}), $U(1)_\chi$ is broken and stable cosmic strings are produced, and the second breaking yields a topologically stable superheavy magnetic monopole. We list the matter and Higgs fields of $SU(5)\times U(1)_\chi$ in Table~\ref{tab:chiSU5}.
%

%
%
\begin{table}[h!]
 \centering
 \begin{tabular}{c | c c c | c c c c }
  \hline \hline
   & \multicolumn{3}{|c|}{Matter Sector} & \multicolumn{4}{|c}{Higgs Sector}\\
   \hline \hline
   &$f$ & $F$ & $\nu^c$ & $h$ & $\Phi$ & $\Psi$  & $S$ \\
   \hline 
     $ SU(5) $  & $\mathbf{\bar 5}_{F}$ & $\mathbf{10}_{F}$  & $\mathbf{1}_F $  & $\mathbf{\bar 5}_{H}$ & $\mathbf{1}_{H}$ & $\mathbf{24}_{H}$ & $\mathbf{1}_H $  \\
   \hline 
   $U(1)_{\chi}$ & 3  & -1 & -5 & -2 & 5 & 0  & 0 \\
   \hline  \hline
  \end{tabular}
 \caption{{Matter and Higgs fields representations of $SU(5)\times U(1)_\chi$ including their respective charges under $U(1)_\chi$.}}
 \label{tab:chiSU5}
\end{table}
The scalar Higgs potential that is invariant under $SU(5)\times U(1)_\chi $ is given by
\bea\label{eq:pot_tot}
V &\supset & V_0 - \mu_\Phi^2 \, |\Phi|^2  + \dfrac{\lambda_1}{4} \, |\Phi|^4 
- {\mu_\Psi^2} \, tr(\Psi^2) - \dfrac{\mu}{3} \, tr(\Psi^3) + \dfrac{\lambda_2}{4} \, tr(\Psi^4)  + \dfrac{\lambda_3}{4} \, \left[tr(\Psi^2)\right]^2  \nonumber \\ 
&& +  \lambda_4 \, |\Phi|^2  \, tr(\Psi^2)  +\dfrac{m^2}{2} S^2 
+  \lambda_5 S^2 \, tr(\Psi^2)  -  \lambda_6 S^2 \, |\Phi|^2    \, ,  
\eea
where $S$ is a real $SU(5)$ singlet scalar field that plays the role of the inflaton, and the $SU(5)$ adjoint representation $\Psi^\alpha_{\,\, \beta}\equiv \psi_a(T^a)^\alpha_{\, \beta}$, with $T^a$ being the $SU(5)$ generators, $a,b,c, \cdots=1,2,\cdots 24$, and $\alpha, \beta, \cdots=1,2,\cdots 5$.
We assume that all dimensionless coefficients in Eq.~(\ref{eq:pot_tot}), are real, and, for simplicity, we will set the coefficient $ \mu=0$.
 The value of the constant vacuum energy term $V_0$ is chosen such that the potential is zero at the true minimum. 
 
For suitably large values of $S$, the 24-plet Higgs field $\Psi$ plays the role of the waterfall field that is frozen at the origin at the start of inflation, while the other Higgs field $\Phi$ follows a field dependent minimum during inflation. The inflaton filed $S$ evolves until it reaches a critical value $S_c$. At this time the waterfall phase transition in hybrid inflation is triggered \cite{Linde:1993cn,Lazarides:2023rqf}, where both $\Psi$ and $\Phi$ evolve towards their true minima at $\langle \Psi \rangle$ and $\langle \Phi \rangle$ respectively \cite{Lazarides:2023rqf}. 
In other words, $U(1)_\chi $ is broken during and after inflation by the non-zero value of $\phi={\rm Re}(\Phi)$, and hence the stable cosmic strings are inflated away. On the other hand the breaking of $SU(5)$ to the SM gauge group occurs after $S$ reaches $S_c$, caused by the 24-plet higgs vev in the SM neutral direction denoted by $ \psi \equiv \psi_{24}$. We denote  the vevs $ \langle \phi \rangle={v_\phi} $, and $ \langle \psi \rangle={v_\psi} $, therefore
\be 
\langle \Psi \rangle=  \dfrac{v_\psi}{\sqrt{15}} {\rm diag}(1,1,1,-3/2,-3/2).
\ee
%

\section{Inflationary potential and classical fields dynamics}
\label{sec:infdynamic}
 The inflationary scenario is driven by the three real scalars $S, \phi$ and $\psi$, with the remaining components of $\Phi$ and $\Psi$ are fixed at zero during and after inflation and do not perturb the inflation dynamics. The inflationary potential then has the form
\bea\label{eq:pot_inf}
V_{\rm inf} &=& V_0 - \dfrac{m_\phi^2}{2} \, \phi^2  + \dfrac{\beta_\phi}{4} \, \phi^4
- \dfrac{m_\psi^2}{2} \, \psi^2 + \dfrac{\beta_\psi}{4} \, \psi^4  + \frac{\beta_{\psi\phi}}{2} \, \psi^2\phi^2   \nonumber \\
&& +\dfrac{m^2}{2} S^2  +  \frac{\beta_{S\psi}}{2} \, S^2 \, \psi^2  -  \frac{\beta_{S\phi}}{2} \,  S^2 \, \phi^2  \, ,
\eea
where the parameters $m_\phi, m_\psi, \beta_\phi, \beta_\psi,\beta_{\psi\phi} , \beta_{S\phi}, \beta_{S\psi}$ are given in terms of the parameters of the original potential in Eq.~(\ref{eq:pot_tot}). 
%
 %
\begin{figure}[htbp!]
    \centering
    \includegraphics[width=0.6\linewidth]{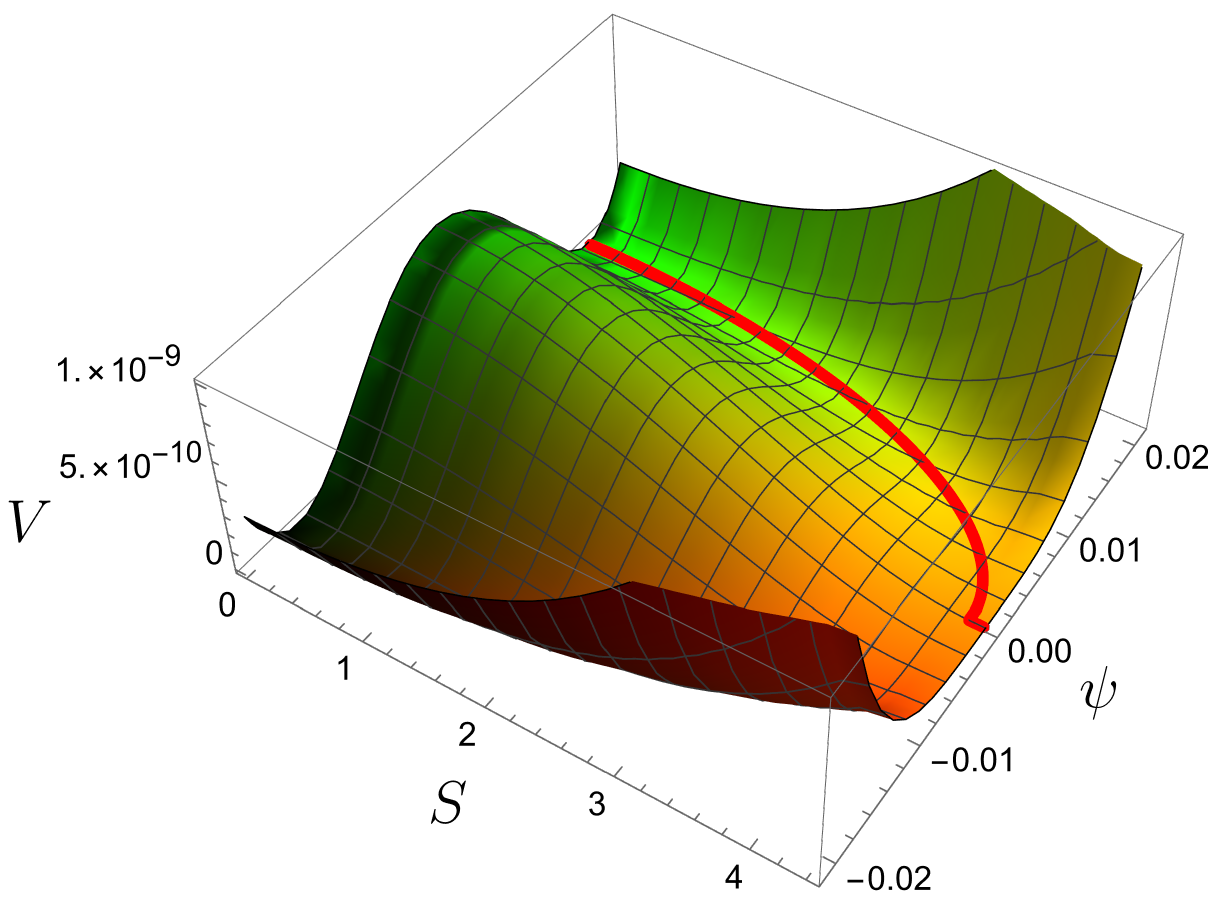}
     \includegraphics[width=0.6\linewidth]{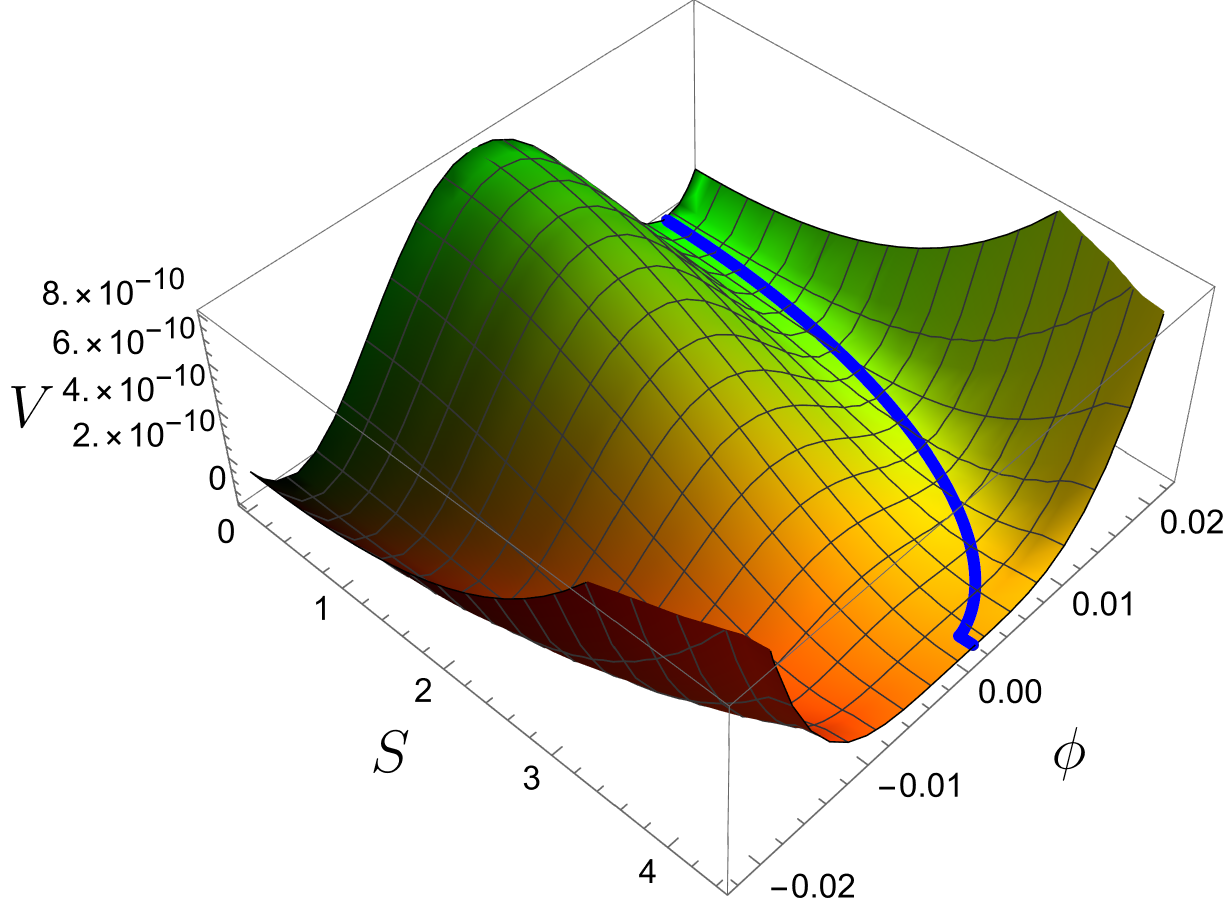}
     \includegraphics[width=0.6\linewidth]{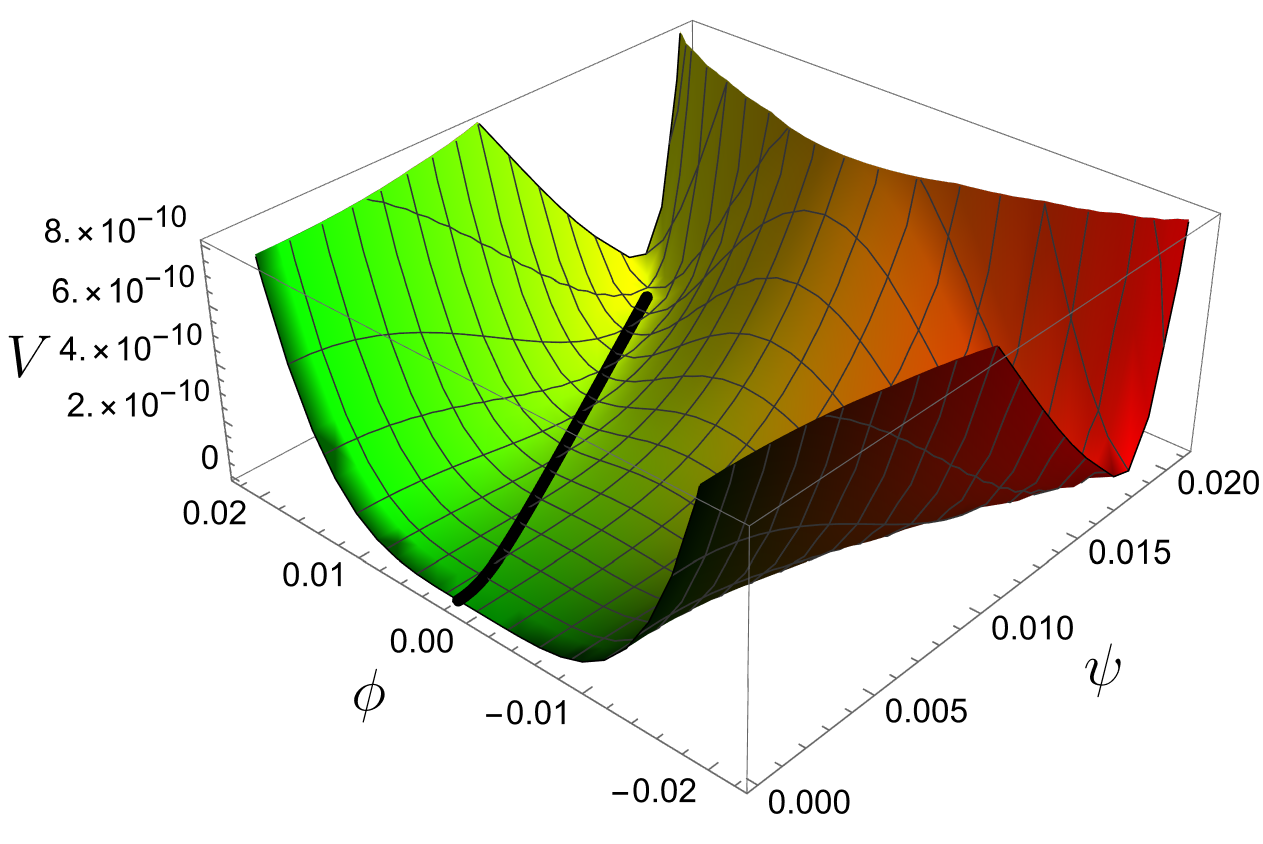}
    \caption{\label{fig:3Dpot}\it The 3D plot of the potential surfaces $V(S,\psi),\, V(S,\phi)$ and $V(\psi,\phi)$. The solid red, blue and black curves represent the trajectories during inflation.}
\end{figure}
At the true minimum of the potential, the vevs of the scalar fields $S$, $\phi$ and $\psi$ are respectively given by
 \bea
 v_S=0, \hspace{1cm}  v_\phi= \sqrt{\frac{\beta _{\psi } m_{\phi }^2-{m_\psi}^2 \beta _{\psi \phi }}{\beta _{\psi } \beta _{\phi }-\beta _{\psi \phi }^2}} , \hspace{1cm}  v_\psi= \sqrt{\frac{\beta_\phi  m_{\psi }^2-\beta _{\psi \phi } m_{\phi }^2}{\beta _{\psi } \beta _{\phi }-\beta _{\psi \phi }^2}},
 \eea
 with 
  \bea
 V_0= \dfrac{1}{4} \left(m_\psi^2 v_\psi^2 + m^2_\phi v_\phi^2 \right) .
 \eea

The mass squared of the inflaton field $S$ at the true minimum is given by
 \bea
M_S^2= m^2+ \beta_{s\psi } v_\psi^2 -  \beta_{s\phi } v_\phi^2 \,,
\eea
while in the $(\psi,\phi)$ basis, the mass-squared matrix at the true minimum is given by
\bea\label{eq:vac-massmat}
{\cal M}^2= 
\left(
\begin{array}{cc}
 2 \beta _{\psi } v_\psi^2 & 2 \beta _{\psi \phi } v_\psi v_\phi  \\
2 \beta _{\psi \phi } v_\psi v_\phi  & 2 \beta _{\phi }  v_\phi^2  \\
\end{array}
\right).
\eea
\begin{figure}[htbp!]
    \centering
    \includegraphics[width=0.6\linewidth]{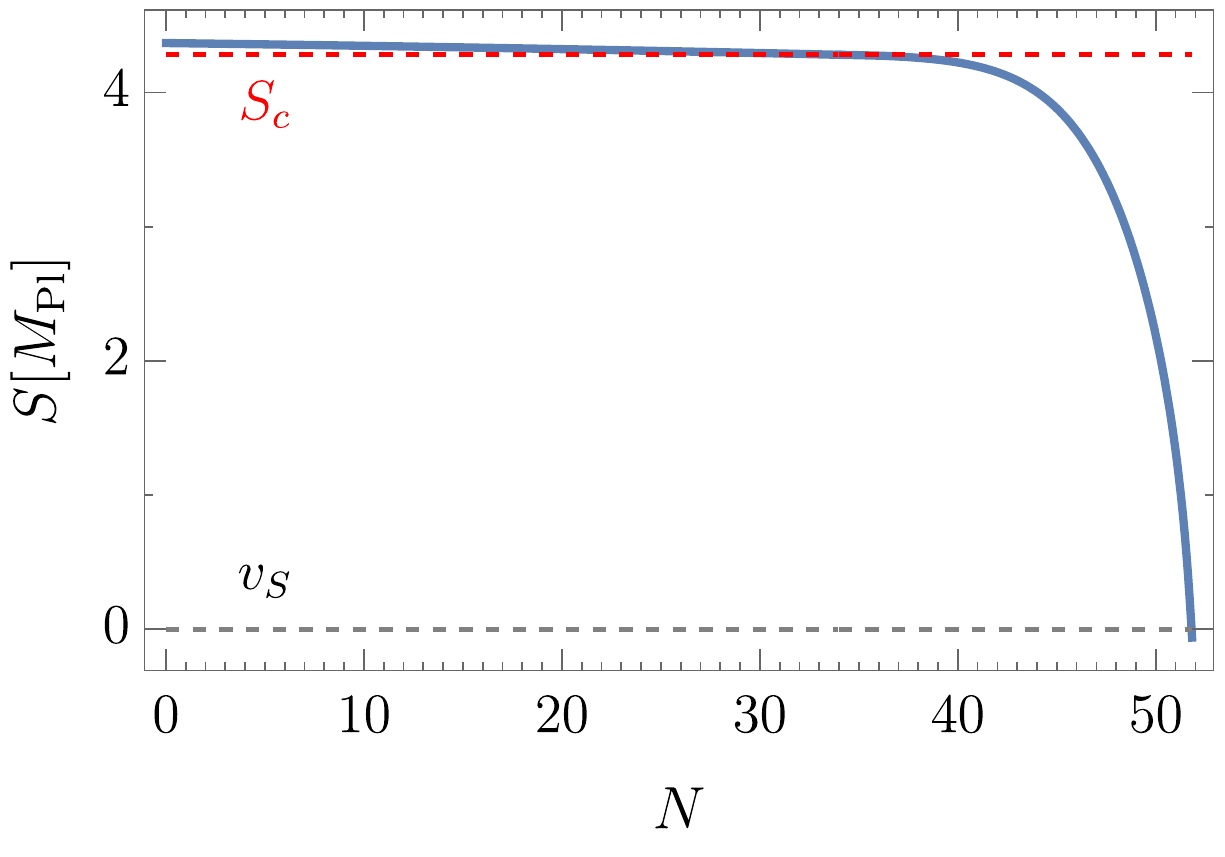}
     \includegraphics[width=0.6\linewidth]{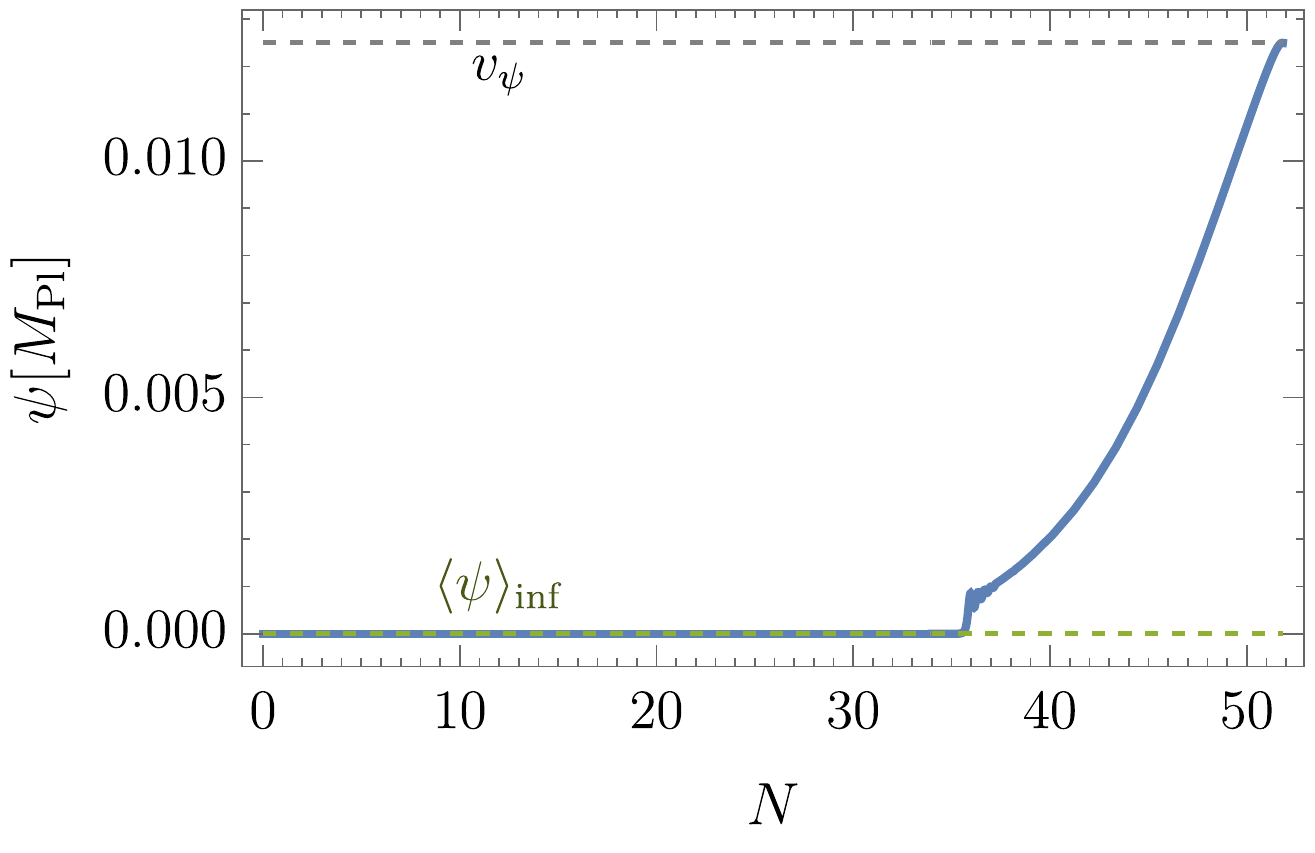}
     \includegraphics[width=0.6\linewidth]{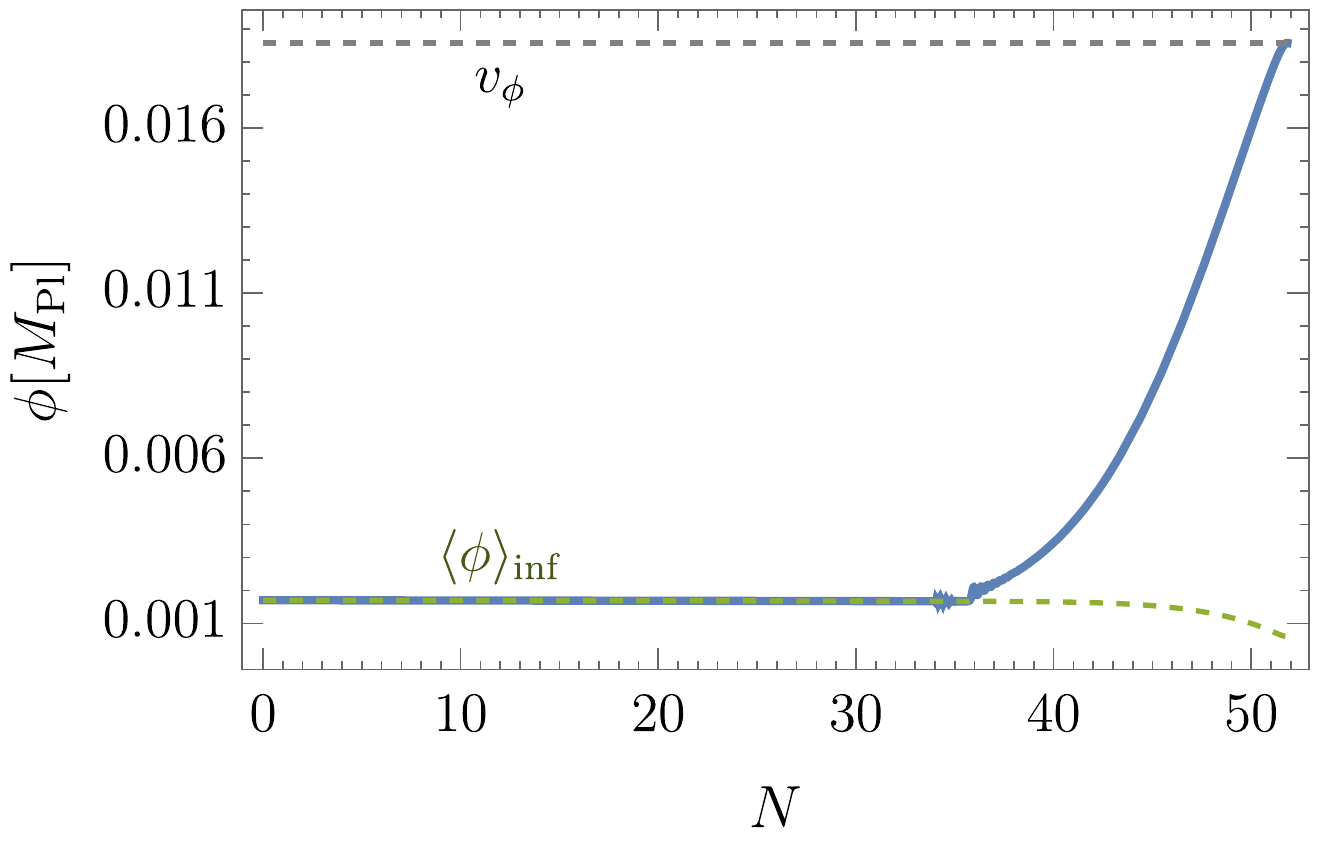}
    \caption{\label{fig:fld-evolution}\it The evolution of the scalar fields (for BP2) during inflation in blue versus the number of e-foldings, as well as their respective vevs at the true minimum in dashed gray. The  green dashed curves represent the values of the waterfall fields that minimize the potential during inflation, before the critical value described by the trajectory~(\ref{eq:trajec}).}
\end{figure}

As advocated in Refs.~\cite{Ibrahim:2022cqs,Lazarides:2023rqf}, the couplings of the inflaton field $S$ to $\Psi$ and $\Phi$ play a crucial role in realizing an inflationary scenario with a hill-top shaped potential, which is a modification of the standard hybrid inflation tree level potential \cite{Linde:1993cn}. Moreover, they control the number of e-foldings after the start of waterfall. The potential in Eq.~(\ref{eq:pot_inf}) is minimized in the $\psi$ and $\phi$ directions, yielding the following trajectory in the $(\psi,\phi)$ plane
\bea\label{eq:trajec} 
(\psi,\phi) = \left(0 \, ,\,  \sqrt{\frac{m_{\phi }^2+  \beta _{S \phi } S^2 }{\beta _{\phi }}} \right) \,.
 \eea
Therefore, with non-zero $\phi$ during inflation, $U(1)_\chi$ is broken and the cosmic strings are inflated away. 
With $\xi= \sqrt{\frac{m_{\phi }^2+  \beta _{S \phi } S^2 }{\beta _{\phi }}}$, the field dependent squared-mass matrix during inflation, in the basis $(S,\phi,\psi)$,  is given by
\bea\label{eq:infmass2}
\!\!\!\!\!\!\!\!\!M^2_{\rm inf}= 
\left(
\begin{array}{ccc}
 m^2- \beta_{S\phi} \, \xi^2        &   \,\,\,\,- 2  \, \beta _{S\phi} \, S \, \xi  & 0 \\
 - 2  \, \beta _{S\phi} \, S \, \xi &    \,\,\,\,2 \beta_\phi \, \xi^2 & 0 \\
 0 & \,\,\,\,0 & -m_{\psi }^2+ \beta _{S\psi}S^2 + \beta _{\psi \phi } \, \xi^2\\
\end{array}
\right). 
\eea
The trajectory~(\ref{eq:trajec}) is valid until  $S$ reaches a critical value $S_c$, where the  mass-squared element $(M^2_{\rm inf})_{\psi\psi}$ flips its sign, with $S_c$ given by
\be 
S_c= \sqrt{\frac{\beta _{\phi } m_{\psi }^2-\beta _{\psi \phi } m_{\phi }^2}{\beta _{\psi \phi } \beta _{{S\phi }}+\beta _{\phi } \beta _{{S\psi }}}} \,.
\ee
 The field $\psi$ stays at the origin as long as $S > S_c$,  where $(M^2_{\rm inf})_{\psi\psi} >0$. As $S$ evolves to smaller values, $\psi$ becomes massless at the moment when $S = S_c$. For $S < S_c$, $(M^2_{\rm inf})_{\psi\psi}<0$, the  waterfall phase is triggered and $SU(5)$ is broken with the production of monopoles. Thus, the trajectory~(\ref{eq:trajec}) becomes unstable, and  $\psi$ and $\phi$ leave the trajectory~(\ref{eq:trajec}) and evolve towards their respective vacuum expectation values at the true minimum of the potential, as depicted in Figure~\ref{fig:fld-evolution}. Figure~\ref{fig:3Dpot} shows the potential surfaces $ V(S,\psi) , V(S,\phi)$ and $ V(\psi,\phi) $, with the solid red, blue and black curves representing the inflation trajectories on the surfaces.
  We are interested in the parameter space where the waterfall field encounters a limited number, on the order of $12$-$22$ $e$-foldings after $S$ crosses the instability point at $S_c$~\cite{Lazarides:2023rqf,Linde:1993cn,Clesse:2010iz, Kodama:2011vs, Clesse:2015wea}. In this case, the observable scales leave the Hubble radius when the fields are still evolving along the trajectory in Eq.~(\ref{eq:trajec}), and  we are able to calculate the inflation observables using the single field slow-roll formalism.

A single field inflation can be realized in the valley in Eq.~(\ref{eq:trajec}), and the tree level effective potential takes the form \cite{Ibrahim:2022cqs,Lazarides:2023rqf}, 
\bea\label{eq:infpot1}
 V_{\text{inf}}(\widetilde{S})= \widetilde{V}_0\left( 1+  \widetilde{S}^2- \gamma \, \widetilde{S}^4\right),
\eea
with the following redefinitions \cite{Ibrahim:2022cqs,Lazarides:2023rqf}
\bea 
 \widetilde{V}_0 \equiv V_0-\dfrac{m_\phi^4}{4\beta_\phi} 
 \,, \hspace{0.5cm} \widetilde{S} \equiv \sqrt{\dfrac{\eta_0}{2}} \, S
 \,, \hspace{0.5cm} \eta_0 \equiv \frac{m^2 \beta _{\phi }-m_{\phi }^2 \beta _{S\phi }}{\widetilde{V}_0 \,\beta _{\phi }} 
 \,, \hspace{0.5cm} \gamma \equiv \frac{\beta _{S \phi }^2}{\eta _0^2 \widetilde{V}_0 \,\beta _{\phi }} \,.
\eea 
Clearly, the potential in~(\ref{eq:infpot1}) has the hilltop shape with a local maximum located at 
\be
\widetilde{S}_{\rm m}=\pm\dfrac{1}{\sqrt{2\gamma}}.
\ee  
The Hubble parameter during inflation is given by $H\approx \sqrt{V_{\text{inf}}/3 M_{\rm Pl}^2}$, where $M_{\rm Pl}=2.42\times 10^{18}$ GeV is the reduced Planck mass.  

In Figure~\ref{fig:fld-evolution}, we present the evolution of the scalar fields during inflation versus the number of $e$-foldings, as well as their respective vevs at the true minimum. The values of the waterfall fields that minimize the scalar potential during inflation, before $S_c$, are denoted by $\langle \psi \rangle_{\rm inf}$ and $\langle \phi \rangle_{\rm inf}$, which trace the trajectory~(\ref{eq:trajec}) depicted by the green dashed curves. The evolution of the scalar fields has a similar behavior for all the benchmark points. We denote the value of $\widetilde{S}$ when the pivot scale $k_*$ exits the inflationary horizon, by $\widetilde{S}_*$. We should satisfy the condition $\widetilde{S}_c < \widetilde{S}_*< \widetilde{S}_{\rm m}$ in order to implement successful hybrid inflation, as the inflaton rolls down to $\widetilde{S}_{c}$ from a value close to $\widetilde{S}_{\rm m}$. 
\begin{figure}[htbp!]
    \centering
    \includegraphics[width=0.8\linewidth]{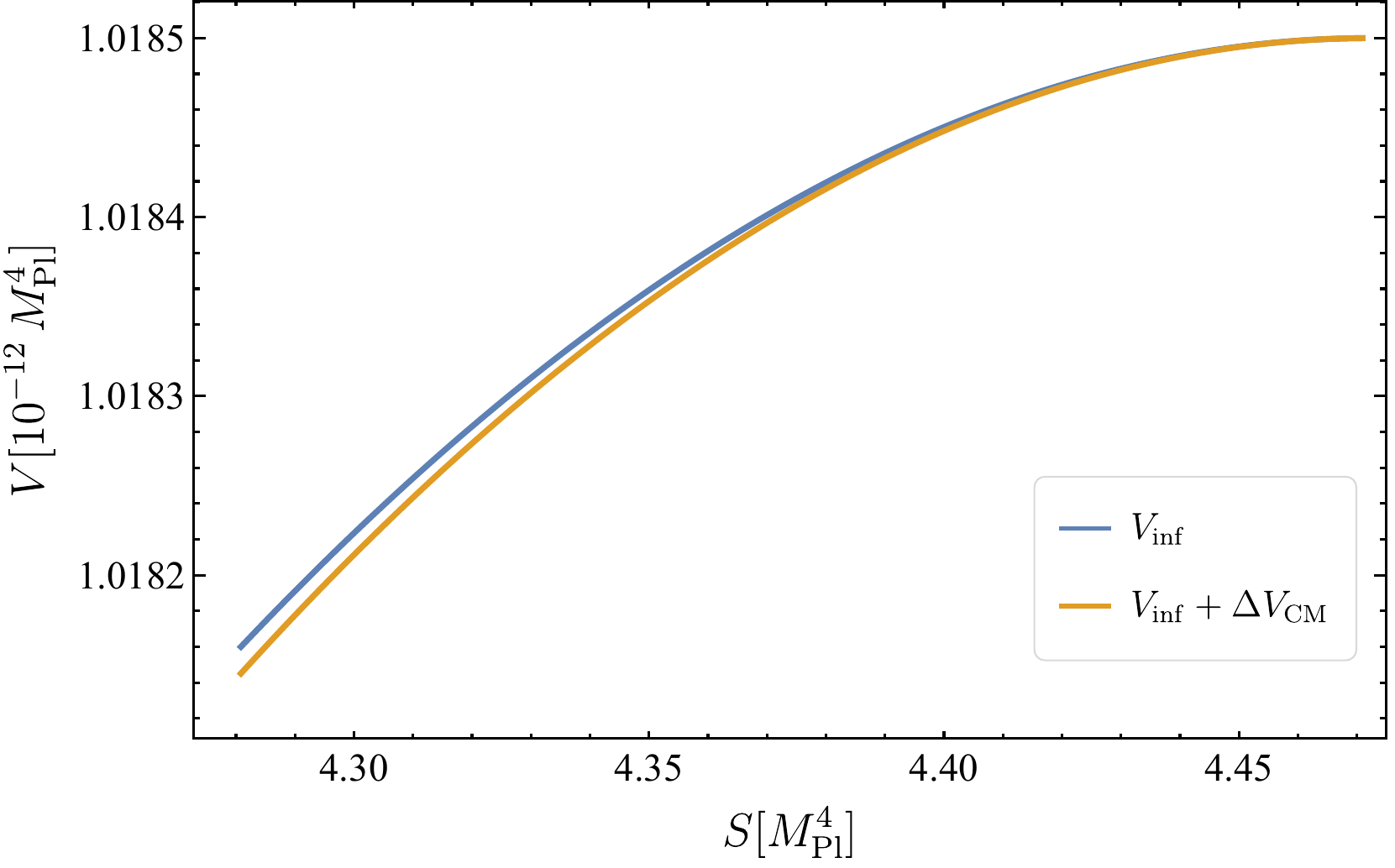}
    \caption{\label{fig:pot}\it The effective single field scalar potential as a function of $S$, between $S_m$ and $S_c$, for the benchmark point BP2. The blue curve represents the tree level potential Eq.~\ref{eq:infpot1}, and the yellow curve represents the total potential including the CW radiative corrections. The CW corrections show a similar behavior for the other benchmark points.}
\end{figure}
\begin{table}[h!]
 \centering
 \begin{tabular}{c |  c  }
  \hline \hline
      Fields            &  Squared masses \\
\hline
\hline
    24 gauge bosons   &     ${5 g^2 \, \xi^2}/{6}   $ \\
    \hline
24 real scalars    &   $ -m_{\psi }^2+ \beta _{S\psi}S^2 + \beta _{\psi \phi } \xi^2$   \\
\hline
1 real scalar   &  $ 2 \beta _\phi \xi^2$  \\
\hline
10 Dirac fermions   &  $ \left(2 Y_S \,S \right)^2 $  \\
  \hline  \hline
  \end{tabular}
 \caption{Squared masses of the scalar fields, vector fields and the vector-like Dirac fermions during inflation for $S\geq S_c$.}
 \label{tab:minfall}
\end{table}

Since the waterfall fields have $S$-dependent masses during inflation, we study the Coleman-Weinberg (CW) radiative corrections  to the tree level inflation potential in Eq.~(\ref{eq:infpot1}). The CW 1-loop correction are given by \cite{Coleman:1973jx} 
\begin{equation}
\Delta V_{\rm CW} = \frac{1}{64 \pi^2} \sum_{i} (-1)^{F_i} M_i^4 \ln\left(M_i^2/\Lambda^2 \right).
\end{equation}
Here $i$ runs over all helicity states, $F_i$ is the fermion number of the $i$th state,
$M_i^2$ denotes the mass squared of the $i$th state along the inflationary path, and $\Lambda$ is a renormalization scale. Table~\ref{tab:minfall} provides the 
relevant squared masses of gauge bosons, $\Phi$, and $\Psi$ components during inflation for $S\geq S_c$. Following Ref.~\cite{Lazarides:2023rqf}, we introduce an extra vector-like pair of fermions $10_{F_1}(4)$,   
$\overline{10}_{F_2}(-4)$, which contribute to the CW corrections, due to their Yukawa coupling $Y_S\, S \, 10_{F1} \, \overline{10}_{F_2} $. The masses squared during inflation of the ten Dirac fermions are given in Table~\ref{tab:minfall}. We then impose the renormalization conditions such that the CW 
correction and its derivative with respect to $S$ vanish at $S=S_m$ \cite{Lazarides:2023rqf}. For $\Lambda \sim 10^{15}$ GeV and $Y_S\sim 10^{-4}$, the CW radiative corrections provide a very small contribution for all benchmark points, as depicted in Figure~\ref{fig:pot}, and the inflationary observables get a slight improvement from these corrections~\cite{Lazarides:2023rqf}.
%
\section{Inflationary observables}
\label{sec:observables}
In this section we study the inflationary observables. We set $M_{\rm Pl}=1$ and drop the subscript ``inf'' for simplicity. In terms of $\widetilde{S}$, the slow-roll parameters are given by
\bea
\epsilon=\frac{\eta_0}{4}\left({\frac{V_{\widetilde{S}}}{V}}\right)^2\,, 
\hspace{0.5cm}
\eta=\frac{\eta_0}{2}\left({\frac{V_{\widetilde{S}\widetilde{S}}}{V}}\right).
\eea
 The total number of e-foldings $\Delta N_*$ between the time when the pivot scale $k_*=0.05 \,{\rm Mpc^{-1}}$ exits the horizon and the end of inflation is calculated from the relation
\bea\label{eq:Ns1}
\Delta N_{*}=\sqrt{\frac{2}{\eta_0}}\bigintss_{\widetilde{S}_{e}}^{\widetilde{S}_*}{\frac{d\widetilde{S}}{\sqrt{\epsilon(\widetilde{S})}}} ,
\eea
such that it coincides with the one calculated from the thermal history of the Universe \cite{Liddle:2003as,Chakrabortty:2020otp,Kawai:2023dac}: 
\begin{equation}\label{eq:Ns2}
\Delta N_* \simeq 61.5 + \frac{1}{2} \mathrm{ln} (\rho_*)-\frac{1}{3(1+\omega_r)} \mathrm{ln} (\rho_e) + \left[\frac{1}{3(1+\omega_r)} - \frac{1}{4} \right]\mathrm{ln} (\rho_r) \ ,
\end{equation}
which is required to solve the horizon and flatness problems. Here, $\widetilde{S}_e$ is the $\widetilde{S}$ value at the end of inflation, $\rho_e = V(\widetilde{S}_e)$ is the energy density at the end of inflation,  $\rho_* = V(\widetilde{S}_*)$  is the energy density of the Universe when the pivot scale exits the horizon, and $\rho_r = (\pi^2/30) g_*T_r^4$ is the energy density at the time of reheating. The parameter $w_r$ is the effective equation-of-state parameter from the end of inflation until reheating that we set equal to zero~\cite{Senoguz:2015lba,Chakrabortty:2020otp}.  We discuss below the upper bound on the reheating temperature. For SM spectrum, we take the effective number of massless degrees of freedom at the reheating time $g_*= 106.75$.\footnote{The value of $g_*$ is of course model dependent.}

We compute the scalar spectral index $n_s$, tensor-to-scalar ratio $r$ and the amplitude of scalar perturbations  $A_s$ at the horizon exit of the pivot scale as follows:
\bea 
n_s=1- 6 \epsilon_* + 2 \eta_* \,, \hspace{0.5cm}  r=16 \epsilon_* \,,  \hspace{0.5cm}   A_s=\frac{V_*}{24 \pi^2 \epsilon_*} .
\eea
\begin{table}[h!]
 \centering
 \begin{tabular}{c |  c c c  }
  \hline \hline
   & $\widetilde{V}_0[M_{\rm Pl}^4]$ & $\eta_0[M_{\rm Pl}^{-2}]$ & $\gamma$    \\
   \hline 
   BP1 &  $0.976 \times 10^{-12}$ & $0.01$ & $7$   \\
 \hline 
   BP2 &  $0.970 \times 10^{-12}$ & $0.01$ & $5$   \\
   \hline 
   BP3 &  $0.980\times 10^{-12}$ & $0.01$ & $4$   \\
  \hline  \hline
  \end{tabular}
 \caption{{ Parameter values for the inflation potential in Eq.~(\ref{eq:infpot1}).}}
 \label{tab:par1}
\end{table}
\begin{table}[h!]
 \centering
 \scalebox{0.9}{
\begin{tabular}{c |  c c c c c c c c}
  \hline \hline
   & $m$ & $m_\psi$ & $m_\phi$  & $\beta_\psi$ & $\beta_\phi$  & $\beta_{\psi\phi}$ & $\beta_{S\psi}$ &  $\beta_{S\phi}$\\
   \hline 
   {BP1} &  $2.53\times 10^{11}$ & $ 7.08\times 10^{13}$ & $2.41\times 10^{14}$  & 0.25 & 0.042 & $-0.102$  & $1.5\times 10^{-8}$  & $5.4\times 10^{-9}$ \\
  \hline  
   {BP2} &  $2.53\times 10^{11}$ & $ 9.57\times 10^{13}$ & $2.41\times 10^{14}$  & 0.042 & 0.028 & $-0.034$  & $5.3\times 10^{-9}$  & $3.7\times 10^{-9}$ \\
   \hline
   {BP3} &  $2.53\times 10^{11}$ & $ 5.5\times 10^{13}$ & $2.41\times 10^{14}$  & 0.054 & 0.026 & $-0.037$  & $5.23\times 10^{-9}$  & $3.2\times 10^{-9}$ \\
   \hline  \hline
  \end{tabular}
  }
 \caption{ Dimensionful parameters, in GeV, and the dimensionless parameters of the potential in Eq.~(\ref{eq:pot_inf}).}
 \label{tab:par2}
\end{table}
\begin{table}[h!]
 \centering
 \begin{tabular}{c |  c c c c c }
  \hline \hline
   & $v_\phi$ & $v_\psi$ & $M_{\psi'}$ & $M_{\phi'}$ & $M_S$  \\
   \hline 
   Obs(BP1) &  $4.66\times 10^{16}$ & $ 3\times 10^{16}$ & $2.888\times 10^{14}$  & $2.3 \times 10^{16}$ & $ 1.134 \times 10^{12}$  \\
   \hline 
   Obs(BP2) &  $4.455\times 10^{16}$ & $ 4\times 10^{16}$ & $2.68\times 10^{14}$  & $1.56 \times 10^{16}$ & $ 1.13 \times 10^{12}$  \\
\hline 
   Obs(BP3) &  $4.67\times 10^{16}$ & $ 3.9\times 10^{16}$ & $2.65\times 10^{14}$  & $1.66 \times 10^{16}$ & $ 1.01 \times 10^{12}$  \\
  \hline  \hline
  \end{tabular}
 \caption{The vevs and physical masses given in GeV. $\phi'$ and $\psi'$ are the mass eigenstates after diagonalizing the mass-squared matrix in Eq.~(\ref{eq:vac-massmat}).}
 \label{tab:ob1}
\end{table}
\begin{table}[h!]
 \centering
 \begin{tabular}{c |  c c c c c c }
  \hline \hline
   & $A_s$ &  $n_s$ & $r$   & $\Delta N_*$ & $\Delta N_c$ & $S_*[M_{\rm Pl}]$ \\
   \hline 
   Obs(BP1) &  $2.19\times 10^{-9}$ & $ 0.9635$ & $3\times 10^{-5}$  & $51.76$ &  $12.76$  & 3.6961  \\
   \hline 
   Obs(BP2) &  $2.16\times 10^{-9}$ & $ 0.963$ & $3\times 10^{-5}$  & $51.03$ &  $17.3$  & 4.388  \\
   \hline 
   Obs(BP3) &  $2.1\times 10^{-9}$ & $ 0.9634$ & $3.35\times 10^{-5}$  & $52.2$ &  $20.9$  & 4.911  \\
  \hline  \hline
  \end{tabular}
 \caption{Model predictions for the CMB observables and number of $e$-foldings.}
 \label{tab:ob2}
\end{table}

The quantity $\widetilde{V}_0$ is determined by the observed value of $A_s= (2.099\pm0.101) \times 10^{-9}$~\cite{BICEP:2021xfz,Planck:2018jri}. 
It turns out that a mild waterfall regime of hybrid inflation~\cite{Clesse:2010iz, Kodama:2011vs,Clesse:2015wea}, with a large number of $e$-foldings (greater than $60$), is found if $\widetilde{S}_*\gg 1$~\cite{Lazarides:2023rqf}. The latter case is not consistent with the Planck inflationary observables~\cite{BICEP:2021xfz,Planck:2018jri}. 
 For $\tilde{S}_* \lesssim 1$, we may have both intermediate or prompt waterfall~\cite{Lazarides:2023rqf}. We focus on the parameter space with the waterfall continuing for a limited number of $e$-foldings, $\Delta N_c \sim 12-21$, where $\Delta N_c$ denotes the number of $e$-foldings between the times corresponding to $S_c$ and $S_e$. We list three benchmark points in Tables~\ref{tab:par1} and~\ref{tab:par2}, with the inflation observables consistent with Planck/BICEP~\cite{BICEP:2021xfz,Planck:2018jri} measurements, as shown in Tables~\ref{tab:ob1} and~\ref{tab:ob2}. The predicted value of the tensor to scalar ratio $r\sim 3\times 10^{-5}$, which can be tested in future CMB experiments such as LiteBIRD~\cite{LiteBIRD:2022cnt} and CMB-S4~\cite{Abazajian:2019eic}.

\section{Reheating and leptogenesis}\label{sec:reheat-leptogenesis}
Inflation ends when the slow-roll parameter $\epsilon_H=1$, and the inflaton field $S$ as well as the waterfall fields $\psi$ and $\phi$ experience damped oscillations about their respective minima, and undergo decays that initiate the reheating phase. The scalar fields $S$ and $\psi$ have effective trilinear couplings to the SM Higgs doublet ${\cal H}$, while  $\phi$ has non-renormalizable couplings to the right-handed neutrinos $\nu_i^c$. The relevant terms in the lagrangian for reheating are given by
\bea\label{eq:Lreheat}
{\cal L} &\supset& - \delta_1 S \, {\cal H}^\dagger   {\cal H} \,  - \delta_2 \psi \, {\cal H}^\dagger   {\cal H} \, + \frac{f}{ M_{\rm Pl}}  \Phi^\dagger \Phi\, \bar{\nu}^c\nu^c \nonumber\\
 &=&  - \delta_1 S \, {\cal H}^\dagger   {\cal H} \,   - \delta_2 \left( \cos(\theta)\psi' - \sin(\theta)\phi' \right) \, {\cal H}^\dagger   {\cal H} \, \nonumber\\
 && + \frac{f}{ M_{\rm Pl}}  \left[ \sin(\theta)\psi' + \cos(\theta)\phi' \right]^2\, \bar{\nu}^c\nu^c
 \,,
\eea
where $f$ is a dimensionless coupling, and $\delta_{1,2}$ are dimensionful couplings that satisfy $\delta_1 \lesssim M_S$ and $\delta_2 \lesssim M_\psi$ in order to preserve perturbativity. $\psi'$ and $\phi'$ are the mass eigenstates with the mixing angle $\theta$ given in terms of the entries of the mass-squared matrix~(\ref{eq:vac-massmat}) by
\bea
\tan(2\theta) =\dfrac{2 ({\cal M}^2)_{12} }{  ({\cal M}^2)_{22} -  ({\cal M}^2)_{11}}.
\eea
The first term in the Lagrangian yields a reheating temperature $T_r \lesssim 10^{14}  \left( \delta_1 / M_S \right)$ GeV, and therefore, $T_r \lesssim 10^{12}$ GeV for $( \delta_1 / M_S) \lesssim 1/100$. Considering a suitably large mixing between $\psi$ and $\phi$, the second term of the Lagrangian yields a reheating temperature $T_r \lesssim 10^{12}$ GeV for $( \delta_2 / M_{\psi'}) \lesssim 0.001$. 

The third term in~(\ref{eq:Lreheat}) provides masses to the right-handed neutrinos denoted by $M_R=\dfrac{f \, v_\phi^2}{M_{\rm Pl}}$ and yields Yukawa couplings  $(M_R / v_\phi) \,  \left[ \sin(\theta)\psi' + \cos(\theta)\phi' \right] \, 10_F \, 10_F $.
The oscillating fields decay mainly into SM higgs and right-handed neutrinos, with a total width  $\Gamma=\Gamma_{\cal H}+\Gamma_N$, where
\bea
\Gamma_{\cal H} &=& \Gamma_{S\to {\cal H}^\dagger {\cal H}}+  \Gamma_{\psi'\to {\cal H}^\dagger {\cal H}}+ \Gamma_{\phi'\to {\cal H}^\dagger {\cal H}}= \frac{1}{8\pi} \left( \frac{\delta_1^2 }{M_S } + \frac{\delta_2^2 \cos^2(\theta) }{M_{\psi'}  }  + \frac{ \delta_2^2 \sin^2(\theta) }{M_{\phi'}  } \right) , \nonumber\\
\Gamma_N &=&   \Gamma_{\psi'\to \nu^c  \nu^c }+ \Gamma_{\phi'\to  \nu^c  \nu^c} \nonumber\\
  &=&      \frac{M_R^2 \cos^2(\theta)  \, M_{\phi'}}{8\pi v_\phi^2}   \left[ 1- \dfrac{M_R^2}{M_{\phi'}^2} \right]  + \frac{M_R^2 \sin^2(\theta)  \, M_{\psi'}}{8\pi v_\phi^2}   \left[ 1- \dfrac{M_R^2}{M_{\psi'}^2} \right] .
\eea
The reheating temperature is estimated to be
\bea 
T_r \approx \left( \frac{90}{\pi^2 \, g_*} \right)^{1/4} \sqrt{\Gamma \, M_{\rm Pl}} \,.
\eea

The last term in the Lagrangian~(\ref{eq:Lreheat}) violates lepton number by two units, $\Delta L=2$, and therefore, leptogenesis can be invoked in our scenario. A lepton asymmetry is generated via right-handed neutrino decays, which is partially converted to baryon asymmetry via sphalerons~\cite{Fukugita:1986hr,Luty:1992un,Lazarides:1990huy,Lazarides:1997ik,Hamaguchi:2002vc,Barbieri:1999ma}. The ratio of lepton number to entropy density, $n_L/s$ is given by \cite{Lazarides:1990huy,Hamaguchi:2002vc}
\bea
\frac{n_L}{s}= \frac{3}{2} \left(\frac{T_r}{m_{\rm inf}}\right) \left(\frac{\Gamma_N}{\Gamma}\right) \, \epsilon_1\,,
\eea
where $m_{\rm inf}$ is the mass of the inflaton that decays into right-handed neutrinos, and $\epsilon_1$ is the CP asymmetry factor, generated
during the out-of-equilibrium decay of the lightest right-handed neutrino $N$ with mass $M_R\lesssim m_{\rm inf}/2$ and $M_R > T_r$. Assuming a normal mass hierarchy among the light active neutrinos, $\epsilon_1$ in our non-SUSY model has the form~\cite{Hamaguchi:2002vc}
\be
\epsilon_1 \simeq -\dfrac{3}{16 \pi}  \dfrac{ m_{\nu_3} M_{R}}{ v^2} \delta_{eff}\,,
\ee
where $v$ is the SM higgs field vev, $\delta_{eff}$ is the effective CP-violating phase with $|\delta_{eff}|\lesssim 1$, and $m_{\nu_3}=0.05$ eV is the mass of the heaviest light neutrino. Therefore, the lepton asymmetry is given by
\be\label{eq:lepasym}
\frac{n_L}{s}= 3\times 10^{-10} \left(\frac{T_r}{10^{6} \, {\rm GeV}}\right) \left(\frac{M_R}{ m_{\rm inf} } \right) \left(\frac{\Gamma_N}{\Gamma}\right) \left(\frac{m_{\nu_3}}{0.05 \, {\rm eV}} \right) \delta_{eff}\,.
\ee
The observed value of the lepton asymmetry is estimated as~\cite{Planck:2018vyg} $|n_L/s|\approx (2.67 - 3.02) \times 10^{-10}$. We can then account for the baryon asymmetry via sphaleron processes that convert an initial lepton asymmetry ${n_L}/{s}$ into the observed baryon asymmetry ${n_B}/{s}= -0.35 \,{n_L}/{s}$~\cite{Khlebnikov:1988sr,Harvey:1990qw}. In our analysis we select appropriate values of the various parameters in~(\ref{eq:lepasym}) to account for the observed value.
%
\section{Waterfall dynamics and production of superheavy monopoles}
\label{sec:wfdynamics}
We study the dynamics during the waterfall phase that continues for about 12-21 e-foldings in our case. We denote the inflation sector fields by $\varphi_n$, where $n$ stands for $S,\psi,\phi$. The classical multi-field dynamics are governed by the Friedmann-Lema\^itre equation
\bea\label{eq:FLeq}
H^2=\dfrac{1}{3M^2_{\rm Pl}} \left[\dfrac{1}{2} \sum_{n=1}^{3} \dot{\varphi}_n^2 +V(\varphi_n) \right],
\eea \label{eq:KGt}
as well as the Klein-Gordon equations of the scalar fields, whose kinetic terms are canonical:
\bea
\ddot{\varphi}_n+3H \dot{\varphi}_n+V_n=0.
\eea
Here, $V_n$ is the derivative of the potential with respect to $\varphi_n$ and a dot denotes the derivative with respect
to the cosmic time $t$. Around the critical point the waterfall field $\psi$ is nearly massless, and therefore experiences quantum fluctuations. We assume an initial displacement $\psi_0\sim H$~\cite{Kodama:2011vs,Clesse:2015wea} at the time when $S=S_c$. In terms of the number of $e$-foldings the equations of motion of the classical fields can be written as
\bea\label{eq:Neqm}
\varphi_n''+ \left( 3 -\epsilon_H \right)  \varphi_n' + \left( 3 -\epsilon_H \right)  \dfrac{V_n}{V} = 0\,.
\eea
 Here the prime denotes the derivative with respect to the number of $e$-foldings variable $N$, and the Hubble slow-roll parameters $\epsilon_H,\eta_H $ are defined as follows
\bea 
\epsilon_H &\equiv&  \dfrac{1}{2H^2} \sum_{n=1}^{3} {\dot{\varphi}_n}^2 =
\dfrac{1}{2} \sum_{n=1}^{3} {\varphi_n'}^2\,, \,\,\,\,\,
 \eta_H \equiv  \epsilon_H -\dfrac{\epsilon_H'}{2 \epsilon_H}  \,.
\eea 
%
\subsection{Multi-field perturbations dynamics and primordial power spectrum}
\label{subsec:modes-PS}
We follow the method in~\cite{Ringeval:2007am,Clesse:2013jra,Clesse:2015wea} for our numerical simulation of the perturbations and the calculation of the primordial power spectrum. The perturbed metric is given by 
\begin{equation}
ds^2=a (\tau) ^2 \left[ -(1+2 \Phi_{\text{B}})d \tau^2+  \left(1-2 \Psi_{\text{B}}\right)\delta_{ij} \,  dx^i dx^j \right], 
\label{eq.metr}
\end{equation}
where $\Phi_{\text{B}}$ and $\Psi_{\text{B}}$ are the Bardeen potentials, which are equal in the longitudinal gauge, $a$ is the scale factor, and $\tau$ is the conformal time that is related to the cosmic time $t$ as $d t\equiv a d \tau $. We use the number of e-foldings as the time variable, such that the scalar fields perturbations $\delta \varphi_n$ as well as $\Phi_{\text{B}}$ evolve according to the equations~\cite{Ringeval:2007am,Clesse:2013jra} 
\bea
	\delta \varphi_n^{''}+(3-\epsilon_H)\delta \varphi_n^{'}+\dfrac{1}{H^2} \sum_{m=1}^{3}V_{n m}\delta \varphi_m+\dfrac{k^2}{a^2H^2}\delta \varphi_n &= & 4\Phi_{\text{B}}^{'}\,\varphi'_n-\dfrac{2\,\Phi_{\text{B}}}{H^2}V_n, \\
	\Phi^{''}_{\text{B}}+(7-\epsilon_H)\,\Phi^{'}_{\text{B}}+\left(2\dfrac{V}{H^2}+\dfrac{k^2}{a^2H^2}\right)\Phi_{\text{B}}  & = &  - \dfrac{1}{H^2} \sum_{m=1}^{3} V_m\,\delta \varphi_m,
\eea 
where $k$ denotes the co-moving wave vector. We set the initial conditions when the modes are well inside the horizon where $k\gg a H $, for each $k$-mode, such that the normalized quantum
modes $v_{n,k}\equiv a\, \delta \varphi_n(k,\tau)$ have free field solutions that behave like plane waves. Therefore, the initial conditions at $N=N_{\rm ic}$ are given  as follows \cite{Ringeval:2007am,Clesse:2013jra}:
\bea \label{eq:IC}
	\delta \varphi_{n}(k,N_{\rm ic}) &=& \dfrac{1}{a \,  \sqrt{2 k} }, \\
	\delta \varphi_{n}^{'} (k,N_{\rm ic})  &=& - \dfrac{1}{ a \, \sqrt{2 k} } \left(1+ i \dfrac{k}{a \, H}\right), \\
\Phi_{\text{B}}(k,N_{\rm ic}) &=& \dfrac{1}{2 \left(\epsilon_H - \dfrac{k^2}{a^2 \, H^2 }\right)}
\mathlarger{\mathlarger{\sum}}_{m=1}^{3} \left(  \varphi_m^{'}\delta \varphi_m^{'}+   3 \varphi_m^{'}\delta \varphi_m + \dfrac{1}{H^2} V_m  \, \delta\varphi_m \right) , \\
	\Phi_{\text{B}}^{'}(k,N_{\rm ic}) &=& \mathlarger{\mathlarger{\sum}}_{m=1}^{3}\dfrac{1}{2} \varphi_m^{'}\delta \varphi_m-\Phi_{\text{B}} \, ,
\eea
where the right hand sides are calculated at $N=N_{ic}$ for each $k$-mode. In numerical simulations, we integrate the background dynamics in order to determine $\Delta N_*$. We then integrate again the background and
the perturbation dynamics simultaneously for each $k$-mode, starting from $N_{\rm ic}$ to $N_e$, taking into account that the sub-Hubble modes
behave like plane waves at $N=N_{\rm ic}$. We can then evaluate the scalar power spectrum $P_\zeta(k)$ from the formula \cite{Ringeval:2007am,Clesse:2013jra}
\bea
\label{eq:PR}
	P_\zeta(k)=\dfrac{k^3}{2\pi^2}\left|\Phi_{\rm B}+\dfrac{\sum\limits_{m=1}^{3} \varphi'_m\delta \varphi_m}{\sum\limits_{m=1}^{3}\varphi^{'2}_m}\right|^2.
\eea
The power spectrum is normalized at the pivot scale $k_*=0.05 \, {\rm Mpc}^{-1}$ to satisfy the Planck constraints~\cite{Planck:2018jri}, and the comoving wave vector is related to the number of $e$-foldings via 
\be 
k=k_*\dfrac{H(N)}{H(N_*)} e^{N-N_*}.
\ee

\begin{figure}[htbp!]
    \centering
    \includegraphics[width=0.8\linewidth]{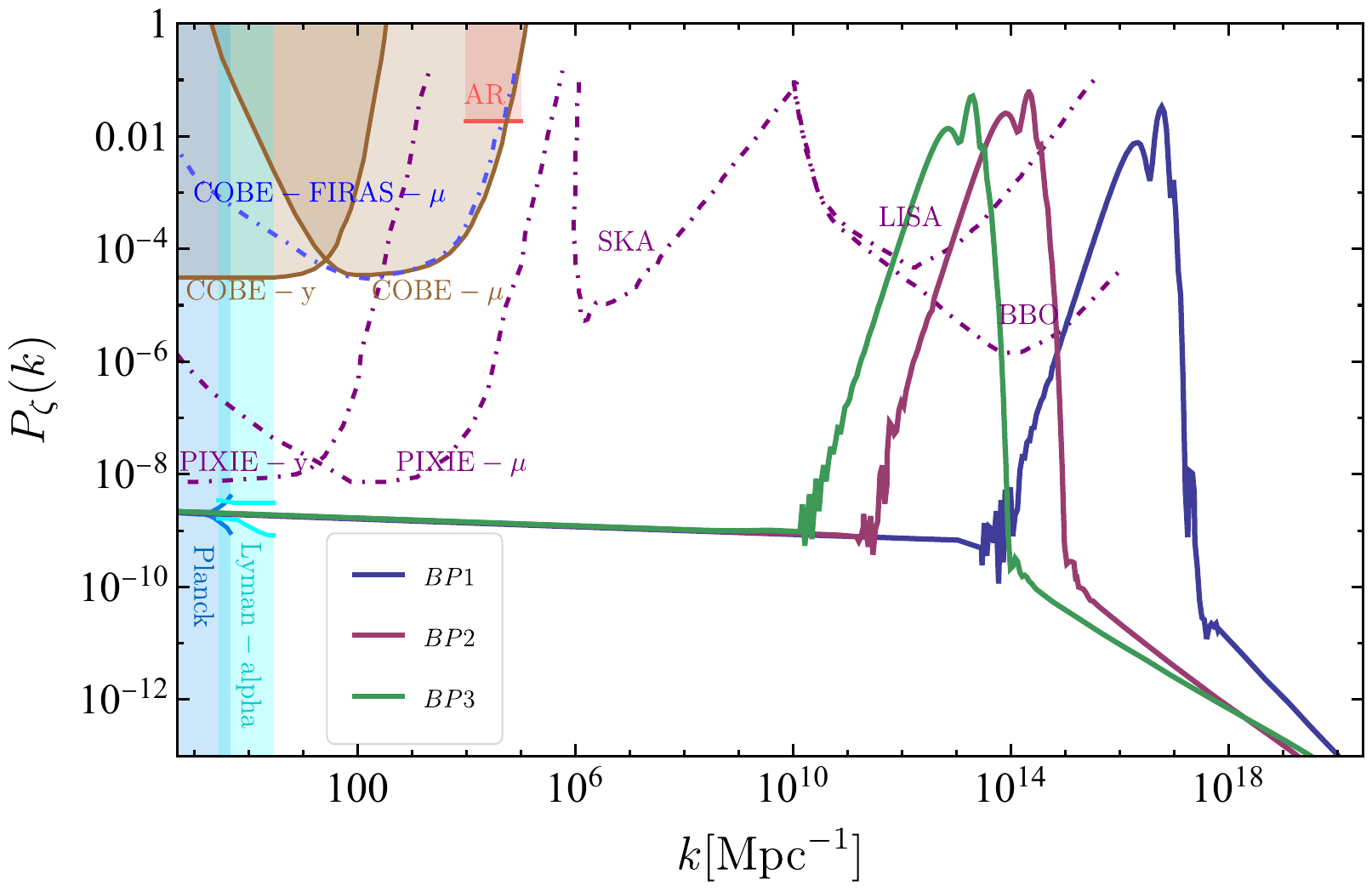}
    \caption{\label{fig:PR}\it The primordial power spectrum, solid curves, corresponding to the three benchmark points in Table~\ref{tab:par1} and~\ref{tab:par2}. The shaded regions display the observational constraints with solid boundaries. Dot-dashed curves represent the expected reach of future experiments.}
\end{figure}

The modes 
 are enhanced due to the phase transition at the tachyonic instability point, and grow exponentially. This induces a peak in the power spectrum with a height $P_\zeta^{\rm peak}$ at the position $k_{\rm peak}$, whose value depends on the Higgs fields vevs $v_\phi$ and $v_\psi$. For $v_\psi\gtrsim 3\times 10^{16}$ and $v_\phi \gtrsim 4\times 10^{16}$ GeV, $P_\zeta^{\rm peak} \gtrsim 10^{-2}$. Therefore we predict that the production of superheavy monopoles at a scale $v_\psi \gtrsim 10^{16}$ GeV is accompanied by a significant peak in the primordial power spectrum. This leads to the formation of primordial black holes and secondary gravitational waves when the curvature perturbations re-enter the horizon during the radiation era, which we discuss in the next section. 

Figure~\ref{fig:PR} shows the curvature power spectrum of the three benchmark points in Table~\ref{tab:par1} and~\ref{tab:par2}. We incorporate the observational constraints on the power spectrum from Planck~\cite{Planck:2018jri}, $\mu$-distortions and $y$-distrortions~\cite{Fixsen:1996nj}, and acoustic reheating (AR)~\cite{Nakama:2014vla}. The dot-dashed curves represent the future PIXIE-like detector exploration of $\mu$-distortions and $y$-distrortions~\cite{Kogut:2011xw} and some future gravitational wave experiments such as LISA, BBO and SKA~\cite{Green:2020jor}. 
It turns out that our model with one waterfall field fixed at the origin and the other shifted from the origin, between $S_*$ and $S_c$, features different shapes of the power spectrum peaks, compared to the standard hybrid inflation with a single waterfall field discussed in~\cite{Clesse:2015wea,Spanos:2021hpk,Braglia:2022phb,Afzal:2024xci}. The latter models have broad-width power spectrum while our model has narrower widths.

For completeness, we discuss the fine-tuning of the power spectrum peak dependence on the symmetry breaking scales $v_\phi$ and $v_\psi$. The fine-tuning can be quantified \cite{Barbieri:1987fn,Leggett:2014mza,Spanos:2021hpk,Afzal:2024xci} by evaluating the quantity
\be 
\Delta_x \equiv {\rm Max} \left|\dfrac{\partial\ln\left(P_\zeta^{\rm Peak} \right)}{\partial \ln(x) }  \right|.
\ee
Since the peak height is sensitive to the values of the vevs, we set $x=v_\phi , \, v_\psi$. The numerical calculations imply that $\Delta_{v_\psi} \sim 10$, and $\Delta_{v_\phi} \sim 100$, which is similar to the result in Ref.~\cite{Spanos:2021hpk}, and is much smaller than the single field inflation value of $\Delta$.
%
\subsection{Superheavy monopole production during the waterfall}
\label{subsec:monopole}

As depicted in Figure~\ref{fig:fld-evolution}, the waterfall starts when $S$ reaches $S_c$. At that time, $\psi$ deviates from the origin and topologically stable monopoles are produced due to $SU(5)$ breaking, with masses about an order of magnitude larger than
the $SU(5)$ symmetry breaking scale \cite{tHooft:1974kcl,Polyakov:1974ek,Lazarides:1980cc,Shafi:1984wk}. The monopoles yield after reheating can be computed from \cite{Maji:2022jzu}   
\bea
Y_M\simeq \frac{45 \, \xi_G^{-3} \,  e^{-3 \, \Delta N_c} }{ 2\pi^2 \,  g_* \,  T_r^3}   \left( \frac{t_e}{t_r}\right)^2 ,
\eea 
where $\xi_G \sim H^{-1}$ is the correlation length, and $t_e$ denotes the time at the end of inflation, $t_c$ is the time when $S=S_c$, and $t_r$ is the time at reheating, which is computed from the relation \cite{Lazarides:1997xr,Lazarides:2001zd}
\be
T_r^2= \sqrt{\dfrac{45}{2\pi^2 \, g_*}} \, \dfrac{M_{\rm Pl}}{t_r}.
\ee 
\begin{table}[h!]
 \centering
 \begin{tabular}{c |  c c c c c   }
 \hline \hline
   & $t_c$ & $t_e$ & $t_r$  & $m_M$[GeV] &  $Y_M$ \\
   \hline 
  Ob( BP1) &  $1.8 \times 10^{-35}$ & $2.49\times 10^{-35}$ & $4.45\times 10^{-28}$  &  $3\times 10^{17}$  & $3.7\times 10^{-28}$  \\
  \hline 
Ob( BP2) &  $1.57 \times 10^{-35}$ & $2.46\times 10^{-35}$ & $9.42\times 10^{-27}$  &  $4\times 10^{17}$  & $0.9\times 10^{-34}$  \\
  \hline 
Ob( BP3) &  $1.446 \times 10^{-35}$ & $2.52\times 10^{-35}$ & $3.7\times 10^{-29}$  &  $3.9\times 10^{17}$  & $3\times 10^{-38}$  \\
  \hline  \hline
  \end{tabular}
 \caption{{Time scales (in second). $m_M$ and $Y_M$ respectively denote the monopole mass and the yield parameter of the monopoles.}}
 \label{tab:parmonopoles}
\end{table}

The observational constraint from the MACRO experiment \cite{Ambrosio:2002qq} on superheavy magnetic monopoles of masses $m_M\gsim 10^{16} $ GeV, provides an upper bound on the monopole flux  $\Phi_M \lesssim 1.4\times 10^{-16}$ cm$^{-2}$ s$^{-1}$ sr$^{-1}$. The latter bound can be translated into an upper bound on the co-moving number density~\cite{Kolb:1990vq}, the yield $Y_M^+=n_M/s \simeq 10^{-27}$, where $s$ is the entropy density and $n_M$ is the monopole number density. We consider a lower bound as a rough threshold for observability of monopoles at $Y_M^{-}=10^{-38}$, that corresponds to monopole flux $ \Phi_M \sim 10^{-27}$ cm$^{-2}$ s$^{-1}$ sr$^{-1}$. We list the monopole yield values predicted by our model in Table~\ref{tab:parmonopoles}. The benchmark point BP1 provides an observable monopole density and scalar induced gravitational waves, but PBHs are evaporated during Big Bang Nucleosynthesis (BBN) and PBHs abundance is strongly constrained. For benchmark points BP2 and BP3, observable monopole density, PBHs as DM and scalar induced gravitational waves are predicted. Interestingly, we can show a link between the observability of monopole fluxes, observable gravitational waves signatures, and PBHs constituting the observed DM abundance in the universe, as we will discuss in the next section.
%

\section{Primordial black holes and gravitational wave signals}
\label{sec:PBH-GW}
After inflation ends, modes with a specific wave number $k$ re-enter the Hubble horizon $H^{-1}$ during the radiation era, and PBHs may be formed  if an overdense region of space-time collapses \cite{Carr:1974nx,Carr:1975qj,Polnarev:1985btg,Garcia-Bellido:1996mdl,Heurtier:2022rhf,Heydari:2023xts}. The enhancement in the power spectrum at the start of the waterfall can be responsible for producing PBHs, if the density contrast $\delta \equiv \delta \rho / \rho$, is larger than a critical value $\delta_c(k)$~\cite{Young:2014ana}.
We assume that the overdensity $\delta$ follows a gaussian centered law, at first order in perturbation theory. Using the Press-Schechter approach, at the time of their formatiom, the PBHs mass fraction $\beta$ compared to the total mass of the Universe can be evaluated from,
\bea
\label{42}
\beta(k)= \frac{1}{\sqrt{2 \pi \sigma ^2 (k)}} \int^{\infty}_{\delta_c(k)} d\delta \,  \exp \left(  -\frac{\delta ^2}{2 \sigma^2(k) } \right) \, .
\eea
Here the implicit $k$-dependence of $\delta_c(k)$ follows from different
epochs during which a $k$ mode re-enters the horizon~\cite{Harada:2013epa}, and $\sigma$ denotes the variance of the curvature perturbations, that can be evaluated from the following formula \cite{Harada:2013epa,Heurtier:2022rhf}
\begin{equation}
\label{40}
\sigma^2 \left(k  \right)= \frac{16}{81}  \int \frac{dk' }{k'} \left(\frac{k'}{k}\right)^4 P_{\zeta}(k') \widetilde{W}\left(\frac{k'}{k}\right).
\end{equation}
 For the window function, we assume a Gaussian distribution function $ \widetilde{W}(x)=e^{-x^2/2} $, and we set $0.4 \lesssim \delta_c \lesssim 0.6$~\cite{Harada:2013epa,Musco:2004ak,Musco:2008hv,Musco:2012au,Musco:2018rwt,Escriva:2019phb,Escriva:2020tak,Musco:2020jjb,Ghoshal:2023wri,Ijaz:2024zma}.\footnote{Other forms of the window function such as a top-hat function~\cite{Young:2019osy,Musco:2020jjb} can be used as well.}
\begin{figure}[htbp!]
    \centering
    \includegraphics[width=0.8\linewidth]{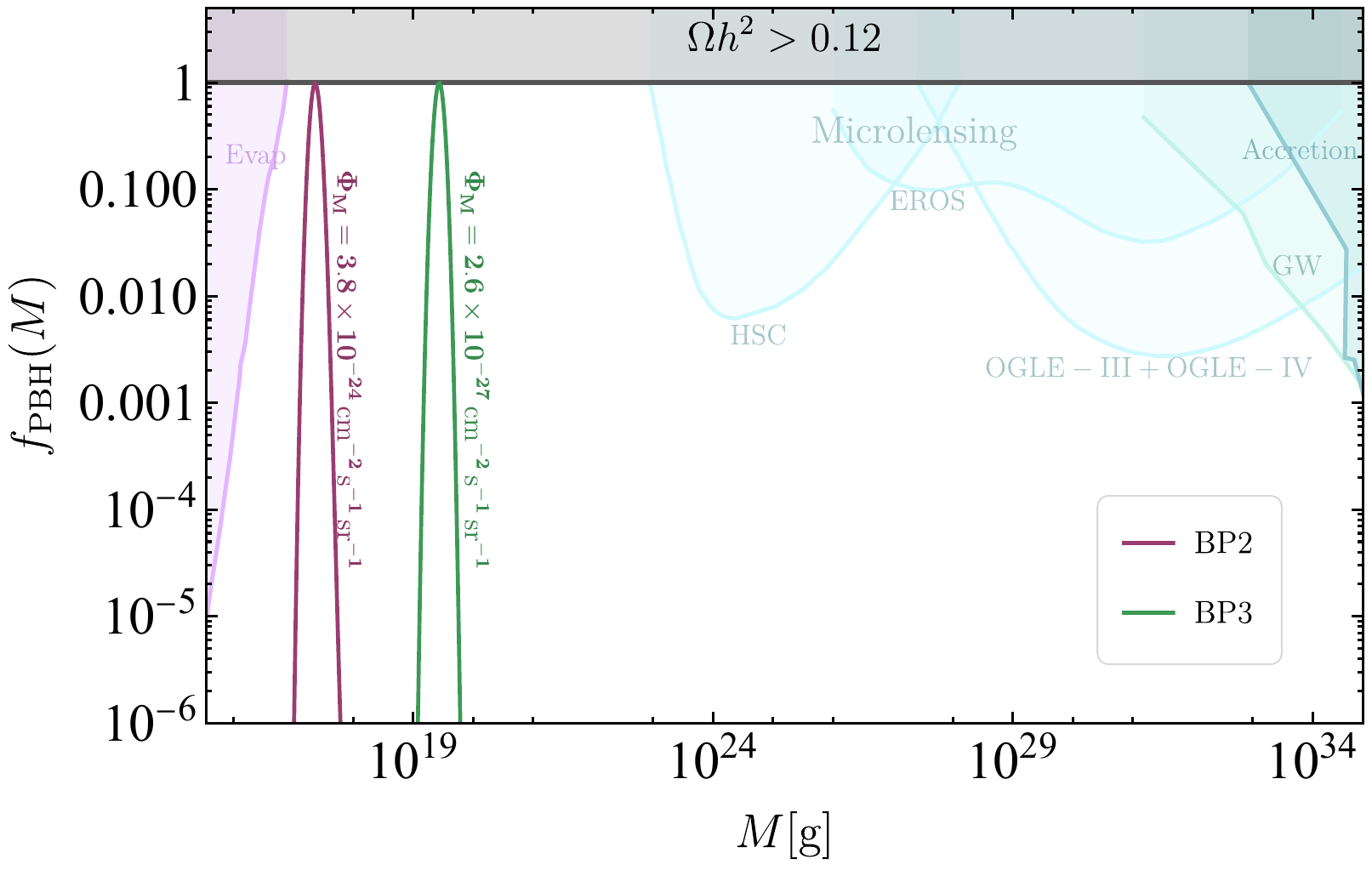}
    \caption{\label{fig:PBH}\it Dark matter abundance of primordial black holes from the waterfall phase transition versus their masses in gram. The shaded regions correspond to the observational constraints.}
\end{figure}
We consider a spherical collapse of perturbations, such that the PBHs mass is given by ~\cite{Ballesteros:2017fsr}
\be
M_{\mathrm{PBH}}=\gamma \frac{4\, \pi \,  \rho}{3}  H^{-3} \,,
\ee
with $\rho$ being the energy density of the universe during collapse to form PBHs. As we consider PBHs creation during the radiation epoch, the PBH mass is given (in grams) in terms of the co-moving wavenumber $k$ \cite{Ballesteros:2017fsr},    
\begin{equation}
\label{eq:MPBHk}
M_{\mathrm{PBH}}(k)=10^{18} \left( \frac{\gamma}{0.2} \right) \left(\frac{g_*(T_f)}{106.75}\right)^{-1/6} \left(\frac{k}{7 \times 10^{13} \, \mathrm{Mpc}^{-1}  }\right)^{-2}  .
\end{equation}
where $T_f$  represents the temperature  at the time of PBHs formation.
 The fractional dark matter abundance of PBHs $f_{\rm PBH}\equiv \Omega_{\rm PBH} / \Omega_{\rm DM}$ can be evaluated from
\begin{equation}
\label{eq:fPBH}
f_{\rm PBH}(M_{\mathrm{PBH}})= 
\frac{\beta(M_{\mathrm{PBH}})}{8 \times 10^{-16}} \left(\frac{\gamma}{0.2}\right)^{3/2} 
       \left(\frac{g_*(T_f)}{106.75}\right)^{-1/4}\left(\frac{M_{\mathrm{PBH}} \,  }{10^{-18 }\; \mathrm{ grams} }\right)^{-1/2}\, , 
\end{equation}
with  $\Omega_{\rm DM} \simeq 0.26$ being the observed DM abundance, and $\gamma\sim 0.2$  is a factor representing the dependence on the gravitation collapse~\cite{Carr:1975qj}. 

In Figure~\ref{fig:PBH}, we display the predicted dark matter abundance of PBHs with masses $M=2.3\times 10^{17}$ g ($1.15\times 10^{-16} \, M_\odot$) and $M=2.7\times 10^{19}$ g ($1.36\times 10^{-14}\, M_\odot$), that is consistent with the various observational constraints, for the benchmark points BP2 and BP3, where the waterfall continues for about 17 and 21 e-foldings respectively after $S=S_c$. For BP1 with about 12.7 efoldings after $S_c$ time, the produced PBHs, with masses $\sim 10^{12}$ g, evaporate at a significant rate during BBN and may change the primordial abundances of light nuclei by emission of energetic particles. Therefore, the PBHs abundance from BP1 is severely constrained and cannot account for the observed DM density~\cite{Carr:2009jm,Carr:2020gox,Keith:2020jww,Thoss:2024hsr,Dienes:2022zgd,Cheek:2022mmy,Barman:2024slw,Haque:2024eyh,Haque:2024cdh}. We have included the various observational constraints on PBHs from black hole evaporation, accretion and GWs~\cite{Carr:2020gox,Green:2020jor,Alexandre:2024nuo,Thoss:2024hsr,Dvali:2020wft,Dvali:2018ytn,Hamaide:2023ayu}, and microlensing, including HSC, EROS and OGLE experiments~\cite{Mroz:2024mse}. The gray region at the top represents a relic abundance that exceeds the observed dark matter density.
%
\begin{figure}[htbp!]
    \centering
    \includegraphics[width=0.8\linewidth]{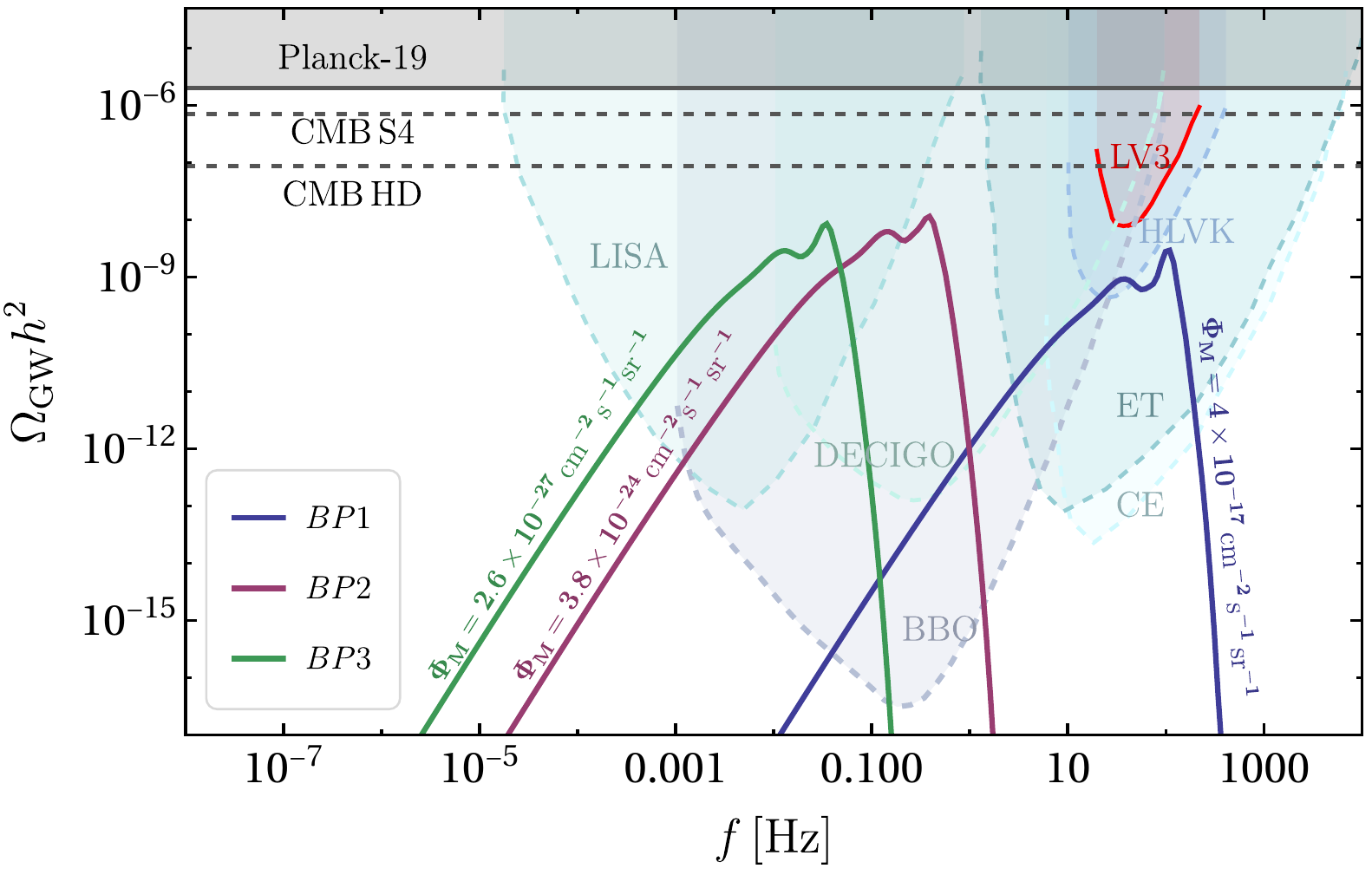}
    \caption{\label{fig:SIGW}\it The  spectrum of scalar induced gravitational waves from the waterfall transition, solid curves, corresponding to the three benchmark points in Table~\ref{tab:par1} and~\ref{tab:par2}. The shaded regions display the observational constraints with solid boundaries, and the expected reach of future GW experiments are shown with dashed boundaries.}
\end{figure}

Next, we discuss the scalar induced gravitational waves associated with the production of PBHs, where the enhanced curvature perturbations are the sources of second order tensor perturbations.  We then use the primordial power spectrum  ${\cal P}_\zeta$ to compute the scalar induced GWs spectrum using the formula \cite{Lewicki:2021xku,Kohri:2018awv,Espinosa:2018eve,Inomata:2019yww,Chatterjee:2017hru}
 \be \label{eq:OmegaGW}
\Omega_{\rm GW}^{\rm si}h^2 \approx 4.6\times 10^{-4} \left(\frac{g_{*,s}^{4}g_{*}^{-3}}{100}\right)^{\!-\frac13} \!\int_{-1}^1 {\rm d} x \int_1^\infty {\rm d} y \, \mathcal{P}_\zeta\left(\frac{y-x}{2}k\right) \mathcal{P}_\zeta\left(\frac{x+y}{2}k\right) F(x,y) \bigg|_{k = 2\pi f} \,.
\ee
Here we set the effective number of entropy degrees of freedom $g_{*,s}\approx g_{*}$ and assume that the gravitational waves were produced during the radiation-dominated epoch. The function $F$ is defined as  
\bea
F(x,y) &=&\frac{(x^2\!+\!y^2\!-\!6)^2(x^2-1)^2(y^2-1)^2}{(x-y)^8(x+y)^8} \times \nonumber \\
&&
\!\left\{\left[x^2-y^2+\frac{x^2\!+\!y^2\!-\!6}{2}\ln\left|\frac{y^2-3}{x^2-3}\right|\right]^{\!2} \!+\! \frac{\pi^2(x^2\!+\!y^2\!-\!6)^2}{4}\theta(y-\sqrt{3}) \right\}.
\eea 

Figure~\ref{fig:SIGW} depicts the gravitational wave spectrum predicted by the $\chi SU(5)$ model corresponding to the benchmark points in Table~\ref{tab:par1} and~\ref{tab:par2}, where the monopoles experience about 13-21 $e$-foldings. It shows the advanced LIGO-VIRGO third run bound~\cite{LIGOScientific:2021nrg} and future GW detectors such as HLVK \cite{KAGRA:2013rdx}, ET \cite{Mentasti:2020yyd}, CE \cite{Regimbau:2016ike}, BBO \cite{Crowder:2005nr, Corbin:2005ny}, LISA \cite{Bartolo:2016ami}, DECIGO \cite{Sato_2017} and LISA \cite{Bartolo:2016ami, amaroseoane2017laser}. As advocated in Section~\ref{sec:wfdynamics}, the power spectrum peak value $P_\zeta^{\rm Peak}$ is controlled mainly by the symmetry breaking scales $v_\psi$ and $v_\phi$, and hence there is a correlation between the gravitational wave peak scale and the monopole scale. It turns out that the gravitational waves produced from such a waterfall  dynamics of hybrid inflation can be accompanied with observable monopole flux which can be tested in the future monopole experiments.
%
\section{Conclusions}
\label{sec:conc}

We have shown that in hybrid inflation models based on realistic GUTs it is possible to simultaneously realize an observable number density of magnetic monopoles as well as primordial black holes and stochastic gravitational wave background. In an example based on $SU(5) \times U(1)_\chi$, the $SU(5)$ symmetry breaking waterfall field experiences a limited number of inflationary $e$-foldings that results in the production of topologically stable superheavy magnetic monopoles. The enhanced scalar perturbations during the waterfall phase also yields primordial black holes that can provide the dark matter in the universe. We have also discussed the scalar induced gravitational wave spectrum produced from the waterfall phase transition, and shown that the peak scale is correlated to the monopole scale. Our model is realistic because the presence of $U(1)_\chi$ can be utilized to explain the origin of neutrino masses as well as the observed baryon asymmetry via leptogenesis. In a more elaborate version of this $SU(5) \times U(1)_\chi$ model, as well as models based on $SO(10)$, it should be possible to simultaneously produce primordial monopoles, cosmic strings and black holes with experimentally observable signatures.
\acknowledgments
The authors would like to thank Drs. Anish  Ghoshal and Ioanna Stamou for useful comments on the manuscript.

\bibliographystyle{mystyle}
\bibliography{GUT1.bib}

\end{document}